\documentclass[12pt,]{article}
\usepackage{amsfonts}
\usepackage{amssymb}

\begin{document}

\title{Inner Structure of $Spin^{c}(4)$ Gauge Potential on $4$-Dimensional
Manifolds}
\author{Xin Liu \thanks{
Corresponding author. Electronic address: xinliu@maths.uq.edu.au} $^a$,
Yi-shi Duan$^b$, Wen-li Yang$^{a,c}$ and Yao-zhong Zhang$^a$}
\date{$^a$ {\small {Department of Mathematics, University of Queensland, QLD
4072, Australia} }\\
$^b$ {\small {Institute of Theoretical Physics, Lanzhou University, Lanzhou
730000, PR China}}\\
$^c$ {\small {Institute of Modern Physics, Northwest University, Xian
710069, PR China}}}
\maketitle

\begin{abstract}
The decomposition of $Spin^{c}(4)$ gauge potential in terms of the Dirac $4$%
-spinor is investigated, where an important characterizing equation $\Delta
A_{\mu }=-\lambda A_{\mu }$ has been discovered. Here $\lambda $ is the
vacuum expectation value of the spinor field, $\lambda =\left\Vert \Phi
\right\Vert ^{2}$, and $A_{\mu }$ the twisting $U(1)$ potential. It is found
that when $\lambda $ takes constant values, the characterizing equation
becomes an eigenvalue problem of the Laplacian operator. It provides a
revenue to determine the modulus of the spinor field by using the Laplacian
spectral theory. The above study could be useful in determining the spinor
field and twisting potential in the Seiberg-Witten equations. Moreover,
topological characteristic numbers of instantons in the self-dual sub-space
are also discussed.\bigskip

\noindent PACS number(s): 02.40.-k, 11.25.Tq, 02.40.Vh \newline
Keyword(s): $Spin^{c}(4)$ gauge potential decomposition; $U(1)$
characterizing equation; Seiberg-Witten theory
\end{abstract}

\newpage

\section{Introduction}

Great importance has been attached to four-dimensional manifolds in the
research of topology \cite{Moore}. In 1983 Donaldson used instantons to
prove new theorems and developed topological invariants for $4$-manifolds.
In 1994 Witten proposed the Seiberg-Witten equations in the study of the $%
Spin^{c}(4)$ bundle and gave a new invariant for classifying $4$-manifolds.
In this paper we will use a physical point of view --- gauge potential
decomposition --- to study the gauge fields and topology on $4$-manifolds.

In recent years gauge potential decomposition has established itself a
useful tool in physicists' mathematical arsenal \cite%
{FadGauPotDec,OurGauPotDecom,DuanLiuCTP1,DuanLiuPRD}. Its main
idea is to reparametrize gauge potentials, such that
characteristic classes --- which are originally conveyed by gauge
field strengths --- can be re-expressed in terms of basic fields
on manifolds. These basic fields can be vector fields, spinor
fields, etc. It is known that characteristic classes are
topological invariants causing non-trivial observable topological
effects in physical systems. Such a change of building blocks will
lead to disclosure of hidden geometric aspects of physical
problems, such as topological charges and locations of
excitations, etc., which are difficult to achieve by other means.
This is the significance of gauge potential decomposition,
consistent with the essence of the Poincar\'{e}-Hopf theorem
\cite{Milnor}.

In this paper our starting point is the decomposition of the $Spin^{c}(4)$
gauge potential. The decomposing basic field employed is the Dirac spinor
field. The results could be useful for the discussion of the Seiberg-Witten
(SW) equations. This paper is arranged as follows. In Sect.2 the
decompositions of the $Spin^{c}(4)$ potential is given. In Sect.3, an
important characterizing equation is obtained. It is found that this
equation will become an eigenvalue problem of the Laplacian operator when
the vacuum expectation value of the spinor field takes constant values. This
study could be applied in determining the twisting potential and spinor
field in the SW equations. In Sect.4, we investigate the topological
characteristic numbers of instantons in the self-dual $SU(2)_{+}$ sub-space.
In Sect.5, we conclude the paper by presenting a summary and discussion.

\section{Decomposition of $Spin^{c}(4)$ Gauge Potential}

Let $\mathcal{M}$ be an oriented closed Riemannian $4$-manifold possessing a
$Spin(4)$-structure, $Spin(4)=SU(2)_{+}\otimes SU(2)_{-}$ \cite{Moore}. Let $%
P$ be a principal $Spin^{c}(4)$-bundle on $\mathcal{M}$. $Spin^{c}(4)$ is
obtained by $U(1)$-twisting $Spin(4)$, $Spin^{c}(4)=Spin(4)\otimes L$, where
$L$ is the twisting line bundle. $Spin(4)$ is the double-cover of the group
space of $SO(4)$. Their Lie algebras $\mathfrak{spin}(4)$ and $\mathfrak{so}%
(4)$ are isomorphic.

Consider a general Dirac equation describing massless neutrinos under an
electromagnetic field,%
\begin{equation}
\gamma ^{\mu }\partial _{\mu }\Psi -\gamma ^{\mu }\omega _{\mu }\Psi
-i\gamma ^{\mu }A_{\mu }\Psi =0.  \label{1stSeibWittEqn'}
\end{equation}%
Here $\mu =1,2,3,4$ denotes the base manifold indices. $D_{\mu }\Psi $ is
the covariant derivative for the Dirac spinor field $\Psi $, $D_{\mu
}=\partial _{\mu }-\left( \omega _{\mu }+iA_{\mu }\right) $. $\left( \omega
_{\mu }+iA_{\mu }\right) $ is the $Spin^{c}(4)$ gauge potential. $A_{\mu }$
is the $U(1)$ potential of $L$, $A_{\mu }\in \mathbb{R}$. $\omega _{\mu }$
is the $SO(4)$ potential, $\omega _{\mu }=\frac{1}{2}\omega _{\mu ab}I_{ab}$%
, with $\omega _{\mu ab}=-\omega _{\mu ba}$. $I_{ab}$ ($a\neq b$) is the $%
SO(4)$ generator realized by the Clifford algebraic $2$-vector, $I_{ab}=%
\frac{1}{4}\left[ \gamma _{a},\gamma _{b}\right] =\frac{1}{2}\gamma
_{a}\gamma _{b}$, with $a,b=1,2,3,4$ denoting the Clifford algebraic
indices. $\gamma ^{\mu }=e_{a}^{\mu }\gamma _{a}$ is the general
Gamma-matrices, raised by the vierbein $e_{a}^{\mu }$ from the Dirac
matrices $\gamma _{a}$.

Our task is to decompose the $Spin^{c}(4)$ potential $\omega _{\mu }+iA_{\mu
}$ in terms of the basic field $\Psi $. Defining $\omega _{abc}=e_{a}^{\mu
}\omega _{\mu bc}$, one has
\begin{equation}
\gamma _{a}\partial _{a}\Psi -\frac{1}{4}\omega _{abc}\gamma _{a}\gamma
_{b}\gamma _{c}\Psi -iA_{a}\gamma _{a}\Psi =0,  \label{3-neutrino}
\end{equation}%
where $\partial _{a}=e_{a}^{\mu }\partial _{\mu }$ and $A_{a}=e_{a}^{\mu
}A_{\mu }$. We notice that $\omega _{abc}$ may be written in three parts:
\begin{equation}
\omega _{abc}=\omega _{abc}^{A}+\omega _{abc}^{S_{1}}+\omega _{abc}^{S_{2}},
\label{3-3partw}
\end{equation}%
where $\omega _{abc}^{A}$ is fully anti-symmetric for $a\;b\;c$, $\omega
_{abc}^{S_{1}}$ symmetric for $a\;b$, and $\omega _{abc}^{S_{2}}$ symmetric
for $a\;c$:
\begin{equation}
\omega _{abc}^{A}=\frac{1}{3}(\omega _{abc}+\omega _{bca}+\omega _{cab}),\
\omega _{abc}^{S_{1}}=\frac{1}{3}(\omega _{abc}+\omega _{bac}),\ \omega
_{abc}^{S_{2}}=\frac{1}{3}(\omega _{abc}+\omega _{cba}).
\end{equation}%
Defining for convenience
\begin{equation}
\bar{\omega}_{b}=2\omega _{aba},\;\;\;\tilde{\omega}_{a}=\epsilon
_{abcd}\omega _{bcd}^{A},
\end{equation}%
one obtains after simple Clifford algebra:
\begin{equation}
\omega _{abc}^{A}\gamma _{a}\gamma _{b}\gamma _{c}=-\gamma _{d}\gamma _{5}%
\tilde{\omega}_{d},\;\;\omega _{abc}^{S_{1}}\gamma _{a}\gamma _{b}\gamma
_{c}=-\frac{1}{3}\bar{\omega}_{c}\gamma _{c},\;\;\omega _{abc}^{S_{2}}\gamma
_{a}\gamma _{b}\gamma _{c}=-\frac{2}{3}\bar{\omega}_{b}\gamma _{b},
\end{equation}%
where $\gamma _{5}=-\gamma _{1}\gamma _{2}\gamma _{3}\gamma _{4}.$ Hence (%
\ref{3-neutrino}) can be rewritten as
\begin{equation}
\gamma _{a}\partial _{a}\Psi +\frac{1}{4}\gamma _{a}(\bar{\omega}_{a}+\gamma
_{5}\tilde{\omega}_{a})\Psi -iA_{a}\gamma _{a}\Psi =0.  \label{4-genDirac3}
\end{equation}

It is easy to see that when the $SO(4)$ gauge potential $\omega _{\mu }$ is
Hermitian, the components $\omega _{\mu ab}$'s are purely imaginary, and
then $\bar{\omega}_{a}$ and $\tilde{\omega}_{a}$ are purely imaginary, $\bar{%
\omega}_{a},\tilde{\omega}_{a}\in i\mathbb{R}$ \cite{ImPotent1}. Thus $\bar{%
\omega}_{a}$ and $\tilde{\omega}_{a}$ are anti-Hermitian and are physical $%
U(1)$-like gauge potentials. Then the conjugate of (\ref{4-genDirac3}) is%
\begin{equation}
\partial _{a}\Psi ^{\dagger }\gamma _{a}-\frac{1}{4}\Psi ^{\dagger }(\bar{%
\omega}_{a}+\gamma _{5}\tilde{\omega}_{a})\gamma _{a}+iA_{a}\Psi ^{\dagger
}\gamma _{a}=0.  \label{4-Dircomconj}
\end{equation}

Eqs.(\ref{4-genDirac3}) and (\ref{4-Dircomconj}) are the starting point for
further decomposing steps. Introduce two real parameters $\rho $ and $\theta
$,
\begin{eqnarray}
\rho &=&\sqrt{(\Psi ^{\dagger }\Psi )^{2}-(\Psi ^{\dagger }\gamma _{5}\Psi
)^{2}},  \label{3-rou} \\
\cosh \theta &=&\frac{\Psi ^{\dagger }\Psi }{\rho },\;\sinh \theta =\frac{%
\Psi ^{\dagger }\gamma _{5}\Psi }{\rho }.
\end{eqnarray}%
One has the so-called duality rotation $e^{\gamma ^{5}\theta }=\cosh \theta
+\gamma ^{5}\sinh \theta $. Then, considering\newline
(A) $\Psi ^{\dagger }\gamma _{b}\times $Eq.(\ref{4-genDirac3})$-$ Eq.(\ref%
{4-Dircomconj})$\times \gamma _{b}\Psi $, we have%
\begin{eqnarray}
0 &=&\left[ \left( \Psi ^{\dagger }\partial _{b}\Psi -\partial _{b}\Psi
^{\dagger }\Psi \right) -2\partial _{a}\left( \Psi ^{\dagger }I_{ab}\Psi
\right) \right]  \nonumber \\
&&+\frac{1}{2}\bar{\omega}_{b}\rho \cosh \theta +\frac{1}{2}\tilde{\omega}%
_{b}\rho \sinh \theta -2iA_{b}\rho \cosh \theta ;  \label{4-SO4decom2'}
\end{eqnarray}%
(B) $\Psi ^{\dagger }\gamma _{b}\gamma _{5}\times $Eq.(\ref{4-genDirac3})$-$
Eq.(\ref{4-Dircomconj})$\times \gamma _{5}\gamma _{b}\Psi $, we have%
\begin{eqnarray}
0 &=&\left[ \left( \Psi ^{\dagger }\gamma _{5}\partial _{b}\Psi -\partial
_{b}\Psi ^{\dagger }\gamma _{5}\Psi \right) -2\partial _{a}\left( \Psi
^{\dagger }\gamma _{5}I_{ab}\Psi \right) \right]  \nonumber \\
&&+\frac{1}{2}\bar{\omega}_{b}\rho \sinh \theta +\frac{1}{2}\tilde{\omega}%
_{b}\rho \cosh \theta -2iA_{b}\rho \sinh \theta .  \label{4-SO4decom3'}
\end{eqnarray}%
Eqs.(\ref{4-SO4decom2'}) and (\ref{4-SO4decom3'}) yield a decomposition for $%
\omega _{\mu }+iA_{\mu }$ \cite{Campolattaro}:%
\begin{eqnarray}
&&\gamma ^{\mu }\left( \omega _{\mu }+iA_{\mu }\right) \Psi  \nonumber \\
&=&-\frac{1}{\rho }\gamma ^{\mu }e^{-\gamma _{5}\theta }e_{\mu b}\left[
\frac{1}{2}\left( \partial _{b}\Psi ^{\dagger }\Psi -\Psi ^{\dagger
}\partial _{b}\Psi \right) +\partial _{a}\left( \Psi ^{\dagger }I_{ab}\Psi
\right) \right.  \nonumber \\
&&\left. +\frac{1}{2}\gamma _{5}\left( \partial _{b}\Psi ^{\dagger }\gamma
_{5}\Psi -\Psi ^{\dagger }\gamma _{5}\partial _{b}\Psi \right) +\partial
_{a}\left( \gamma _{5}\Psi ^{\dagger }\gamma _{5}I_{ab}\Psi \right) \right]
\Psi .  \label{4-SO(4)finaldecomp66}
\end{eqnarray}

The expression (\ref{4-SO(4)finaldecomp66}) is generally covariant for the
coordinates of the Riemannian base manifold. One should eliminate from (\ref%
{4-SO(4)finaldecomp66}) the arbitrariness in the evaluation of the metric.
Following the routine of general relativity a coordinate condition for the
vierbein $e_{\mu b}$ should be adopted as a restraint. We choose the
harmonic coordinate condition \cite{harmcoorcond}:
\begin{equation}
\frac{1}{\sqrt{g}}\partial _{\nu }\left( \sqrt{g}g^{\nu \lambda }\right) =0\
\ \ i.e.\ \ \ e_{a}^{\mu }\partial _{\mu }e_{a}^{\nu }+\frac{1}{2}\omega
_{a}e_{a}^{\nu }=0.  \label{4-harmonicoord}
\end{equation}%
Under this condition the coordinate system will degenerate to the inertial
coordinate system when the gravity vanishes \cite{Fok}. Defining $\gamma
^{\mu }{}_{\nu }=e_{a}^{\mu }e_{\nu b}I_{ab}$, we arrive at the Dirac spinor
decomposition expression for the $Spin^{c}(4)$ gauge potential:%
\begin{eqnarray}
&&\gamma ^{\mu }\left( \omega _{\mu }+iA_{\mu }\right) \Psi   \nonumber \\
&=&\gamma ^{\mu }\frac{e^{-\gamma _{5}\theta }}{\rho }\left\{ \frac{1}{2}%
\left[ \Psi ^{\dagger }\partial _{\mu }\Psi -(\partial _{\mu }\Psi ^{\dagger
})\Psi \right] +\frac{\gamma _{5}}{2}\left[ \Psi ^{\dagger }\gamma
_{5}\partial _{\mu }\Psi -(\partial _{\mu }\Psi ^{\dagger })\gamma _{5}\Psi %
\right] \right.   \nonumber \\
&&\left. -\nabla _{\nu }\left( \Psi ^{\dagger }\gamma ^{\nu }{}_{\mu }\Psi
\right) -\nabla _{\nu }\left( \gamma _{5}\Psi ^{\dagger }\gamma ^{\nu
}{}_{\mu }\gamma _{5}\Psi \right) \right\} \Psi .
\label{4-SO(4)finaldecomp7}
\end{eqnarray}%
It is recognized immediately that the first two terms in (\ref%
{4-SO(4)finaldecomp7})
\begin{equation}
j_{\mu }=\frac{1}{2}\left[ (\partial _{\mu }\Psi ^{\dagger })\Psi -\Psi
^{\dagger }\partial _{\mu }\Psi \right] ,\;\;\;\tilde{j}_{\mu }=\frac{1}{2}%
\left[ (\partial _{\mu }\Psi ^{\dagger })\gamma _{5}\Psi -\Psi ^{\dagger
}\gamma _{5}\partial _{\mu }\Psi \right]
\end{equation}%
are respectively the velocity current and pseudo-velocity current in quantum
mechanics. For the last two terms, recall the covariant first Maxwell
equation in the generally covariant $U(1)$ electromagnetics \cite%
{Campolattaro}:
\begin{equation}
\nabla _{\nu }(f^{\nu }{}_{\mu })=J_{\mu },
\end{equation}%
where $f^{\nu }{}_{\mu }$ denotes a field strength tensor and $J_{\mu }$ an
electromagnetic current. It is recognized that $\left( \Psi ^{\dagger
}\gamma ^{\nu }{}_{\mu }\Psi \right) $ and $\left( \gamma _{5}\Psi ^{\dagger
}\gamma ^{\nu }{}_{\mu }\gamma _{5}\Psi \right) $ are two $U(1)$ field
tensors
\begin{equation}
F^{\nu }{}_{\mu }=\Psi ^{\dagger }\gamma ^{\nu }{}_{\mu }\Psi ,\;\;\;\tilde{F%
}^{\nu }{}_{\mu }=\gamma _{5}\Psi ^{\dagger }\gamma ^{\nu }{}_{\mu }\gamma
_{5}\Psi .  \label{4-SeibWitttensor}
\end{equation}%
Thus the last two terms of (\ref{4-SO(4)finaldecomp7}) become two currents:
\begin{equation}
\nabla _{\nu }(\Psi ^{\dagger }\gamma ^{\nu }{}_{\mu }\Psi )=J_{\mu
},\;\;\;\nabla _{\nu }(\Psi ^{\dagger }\gamma ^{\nu }{}_{\mu }\gamma
_{5}\Psi )=\tilde{J}_{\mu }.  \label{4-SeibWittCurrent}
\end{equation}%
And (\ref{4-SO(4)finaldecomp7}) reads
\begin{equation}
\gamma ^{\mu }\left( \omega _{\mu }+iA_{\mu }\right) \Psi =-\frac{1}{\rho }%
\gamma ^{\mu }e^{-\gamma _{5}\theta }\left[ j_{\mu }+\gamma _{5}\tilde{j}%
_{\mu }+J_{\mu }+\gamma _{5}\tilde{J}_{\mu }\right] \Psi .  \label{decomp1}
\end{equation}%
In the following sections the essence and significance of $J_{\mu }$ and $%
\tilde{J}_{\mu }$ will be further revealed.

The above Dirac spinor decomposition of the $Spin^{c}(4)$ gauge potential,
Eq.(\ref{4-SO(4)finaldecomp7}) i.e. Eq.(\ref{decomp1}), is of independent
importance. It could find applications in many physical problems besides in
the current paper. For instance, when $A_{\mu }=0$, (\ref{decomp1}) gives
the Dirac spinor decomposition of the $SO(4)$ potential, which may be useful
for obtaining topological excitations in higher-dimensional quantum Hall
systems and the Dirac spinor structure of the $SO(4)$ Chern-Simons form \cite%
{DuanLiuPRD,SO(4)C-S}.

\section{Laplacian Characterizing Equation}

Eq.(\ref{decomp1}) can be written as
\begin{equation}
\gamma ^{\mu }\left( \omega _{\mu }+iA_{\mu }\right) \Psi =-\gamma ^{\mu }%
\frac{\Psi ^{\dagger }\Psi -\gamma _{5}\Psi ^{\dagger }\gamma _{5}\Psi }{%
(\Psi ^{\dagger }\Psi )^{2}-(\Psi ^{\dagger }\gamma _{5}\Psi )^{2}}\left[
j_{\mu }+\gamma _{5}\tilde{j}_{\mu }+J_{\mu }+\gamma _{5}\tilde{J}_{\mu }%
\right] \Psi .  \label{4-SpinCdec2}
\end{equation}%
The $Spin^{c}(4)$ space is a direct sum of the self-dual and anti-self-dual
sub-spaces, denoted respectively by $\left( S^{\pm }\otimes L\right) $. The
spinor fields in them are respectively $\Psi _{\pm }$: $\gamma _{5}\Psi
_{\pm }=\pm \Psi _{\pm }$. For the purpose of discussing instantons and the
Seiberg-Witten equations in the following text, we focus on the self-dual $%
S^{+}\otimes L$. Then (\ref{4-SpinCdec2}) becomes%
\begin{eqnarray}
&&\gamma ^{\mu }\left( \omega _{\mu }+iA_{\mu }\right) \Psi  \nonumber \\
&=&-\gamma ^{\mu }\frac{I}{\Psi ^{\dagger }\Psi }\left[ \frac{1}{2}\left(
(\partial _{\mu }\Psi ^{\dagger })\Psi -\Psi ^{\dagger }\partial _{\mu }\Psi
\right) +\nabla _{\nu }(\Psi ^{\dagger }\gamma ^{\nu }{}_{\mu }\Psi )\right]
\Psi ,  \label{4-SpinCdec4}
\end{eqnarray}%
where for notation simplicity \textquotedblleft $\Psi _{+}$%
\textquotedblright\ has been denoted by \textquotedblleft $\Psi $%
\textquotedblright , and $\gamma _{5}$ takes value $+1$.

Choose the Dirac-Pauli representation for the $\gamma $-matrices:%
\begin{equation}
\gamma _{i}=\left(
\begin{array}{ll}
0 & -i\sigma _{i} \\
i\sigma _{i} & 0%
\end{array}%
\right) ,\;\gamma _{4}=\left(
\begin{array}{ll}
I & 0 \\
0 & -I%
\end{array}%
\right) ,\;\gamma _{5}=\left(
\begin{array}{ll}
0 & I \\
I & 0%
\end{array}%
\right) ,
\end{equation}%
with $\sigma _{i},\;i=1,2,3$, being the Pauli matrices. Thus $\Psi =(%
\begin{array}{ll}
\Phi , & \Phi
\end{array}%
)^{T}$, where $\Phi $ is a Pauli $2$-spinor. Then considering the Maxwell
equation in terms of the self-dual $\Psi $ \cite{Campolattaro},
\begin{equation}
\partial _{\mu }A_{\nu }-\partial _{\nu }A_{\mu }=-i\Psi ^{\dagger }\gamma
_{\mu \nu }\Psi ,  \label{2ndSeibWittEqn}
\end{equation}%
(\ref{4-SpinCdec4}) becomes
\begin{equation}
\gamma ^{\mu }\left( \omega _{\mu }+iA_{\mu }\right) \Psi =\gamma ^{\mu }
\left[ \frac{I}{2\Phi ^{\dagger }\Phi }(\Phi ^{\dagger }\partial _{\mu }\Phi
-\partial _{\mu }\Phi ^{\dagger }\Phi )-i\frac{1}{\Phi ^{\dagger }\Phi }%
\Delta A_{\mu }\right] \Psi ,  \label{4-SpinCdec5}
\end{equation}%
where $\Delta =\frac{1}{\sqrt{g}}\partial _{\nu }(g^{\nu \lambda }\sqrt{g}%
\partial _{\lambda })$ is the Laplacian operator, and the Lorentz gauge $%
\nabla _{\nu }A^{\nu }=0$ has been imposed.

In the previous work \cite{DuanLiuCTP1} we have obtained the decomposition
of an $SU(2)$ gauge potential $\varpi =\frac{1}{2i}\varpi _{\mu }^{i}\sigma
_{i}dx^{\mu }$ in terms of the Pauli spinor $\Phi $%
\begin{equation}
\varpi _{\mu }=\mathfrak{j}_{\mu }(\Phi )-\frac{1}{2}Tr\left[ \mathfrak{j}%
_{\mu }(\Phi )\right] I,  \label{SU(2)decomp2}
\end{equation}%
where $\mathfrak{j}_{\mu }(\Phi )=\frac{1}{\Phi ^{\dagger }\Phi }\left(
\partial _{\mu }\Phi \Phi ^{\dagger }-\Phi \partial _{\mu }\Phi ^{\dagger
}\right) $. The $U(1)$ projection of $\varpi _{\mu }$ onto the direction of
spin $s_{i}=\frac{\Phi ^{\dagger }\sigma _{i}\Phi }{2i\Phi ^{\dagger }\Phi }$
is
\begin{equation}
\varpi _{\mu }^{i}s_{i}=\frac{i}{2\Phi ^{\dagger }\Phi }\left[ \Phi
^{\dagger }\partial _{\mu }\Phi -\partial _{\mu }\Phi ^{\dagger }\Phi \right]
.  \label{SU(2)decomp1}
\end{equation}%
Comparing the first term on the RHS of (\ref{4-SpinCdec5}) with (\ref%
{SU(2)decomp1}), one sees that the former gives rise to the $SU(2)_{+}$
gauge potential $\varpi _{\mu }$ in the self-dual sub-space. Therefore, for (%
\ref{4-SpinCdec5}) an important case is discovered: if, in particular, the
following equation holds,%
\begin{equation}
\Delta A_{\mu }=-\lambda A_{\mu },\ \ with\ \ \lambda =\Phi ^{\dagger }\Phi ,
\label{eigenValLap1}
\end{equation}%
then (\ref{4-SpinCdec5}) leads to an $SU(2)$-type decomposition. (\ref%
{eigenValLap1}) can be regarded as a characterizing equation of the twisting
potential $A_{\mu }$, which provides a possible revenue to study $%
Spin^{c}(4) $ gauge field by investigating the $\left( S^{+}\otimes L\right)
$ sub-space.

There are three points to address for (\ref{eigenValLap1}). Firstly, a
remarkable case for (\ref{eigenValLap1}) is that $\lambda $ takes constant
values. This makes the equation become an eigenvalue problem of the
Laplacian operator $\Delta $, with $\lambda $ being the eigenvalue and $%
A_{\mu }$ the eigenfunction. This Laplacian eigenvalue problem has been
investigated by many authors \cite{Yau}. Important theorems on the spectrum $%
\left\{ \lambda _{(k)},f_{(k)};k=0,1,2,\cdots \right\} \;$ have been
obtained, including those concerning the estimates and properties of the
eigenvalues. Actually, if the base manifold has no boundary, $\Delta $ is an
order-$2$ elliptic self-adjoint operator with discrete eigenvalues $%
\{0=\lambda _{(0)}<\lambda _{(1)}\leq \lambda _{(2)}\leq \cdots \}$, while
the non-trivial eigenfunctions $\left\{ f_{(k)}\in C^{\infty }(\mathcal{M}%
)\right\} $ form an orthogonal basis in the Hilbert space. $\left\{ \lambda
_{(k)}\right\} $ is the spectrum of the gauge-invariant vacuum expectation
value of the self-dual spinor field $\Psi $. This provides a method to study
$\left\{ \lambda _{(k)}\right\} $ and the $U(1)$ twisting potential $A_{\mu }
$.

Secondly, for the case of the zero-eigenvalue $\lambda _{(0)}=\Phi ^{\dagger
}\Phi =0$ (i.e. $|\Phi |=0$), the corresponding solution space is
finite-dimensional. This case will be discussed in detail in Sect.4. It will
be shown that the zero-points of $\Phi $ correspond to the instanton
solutions, whose topological charges account for the second Chern numbers.

Thirdly, the Laplacian operator $\Delta $ is an unbounded operator, as $%
\left. \lambda _{(k)}\right\vert _{k\rightarrow \infty }$ $\rightarrow
\infty $. One can introduce a heat operator $e^{t\Delta }$, which is bounded
because $\left. e^{-t\lambda _{(k)}}\right\vert _{k\rightarrow \infty }$ $%
\rightarrow 0$ for $Re\ t>0$. Defining $\mathfrak{F}_{(k)}=e^{t\Delta
}f_{(k)}$ from the above eigenfunctions $\left\{ f_{(k)}\right\} $, it is
easy to check that $\mathfrak{F}_{(k)}$ serves as the solution of the heat
equation $(-\Delta +\partial _{t})\mathfrak{F}=0$. The solution $\mathfrak{F}
$ is given by the so-called heat kernel function $H(\vec{x},\vec{y},t)$%
\begin{equation}
(e^{t\Delta }f)(x)=\int_{\mathcal{M}}H(\vec{x},\vec{y},t)f(\vec{y}%
),\;\;\;\forall f\in L^{2}(\mathcal{M}).
\end{equation}%
The concrete form of $H(\vec{x},\vec{y},t)$ depends on the background
geometry of the base manifold $\mathcal{M}$ \cite{Yau}.

We address that the above theory could be useful for the study of the
Seiberg-Witten (SW) theory. The SW theory determines the exact low-energy
effective Lagrangian of the $\mathcal{N}=2$ supersymmetric Yang-Mills gauge
theory in terms of a single prepotential $\mathcal{F}$ \cite%
{SeibWittPhys,Alv-Gau}. It resulted in a revolution in string theory in
1994, and also played an important role in topology \cite{Moore} by
providing a powerful SW invariant for the classification of four-dimensional
manifolds. Taubes proved that the SW invariants are equivalent to the Gromov
invariants for symplectic manifolds.

Let $\mathfrak{M}$ be an oriented closed Riemannian $4$-manifold possessing
a $Spin(4)$-structure. Usually $\mathfrak{M}$ is considered to be
boundaryless. Let $\mathcal{P}$ be a principal $Spin^{c}(4)$-bundle on $%
\mathfrak{M}$. The SW equations read
\begin{eqnarray}
\mathbf{D}^{+A}\Psi _{+} &=&\gamma ^{\mu }\partial _{\mu }\Psi _{+}-\gamma
^{\mu }\Omega _{\mu }\Psi _{+}-i\gamma ^{\mu }A_{\mu }\Psi _{+}=0,
\label{Seiberg-Witten1real} \\
F_{+\mu \nu } &=&\partial _{\mu }A_{\nu }-\partial _{\nu }A_{\mu }=-i\Psi
_{+}^{\dagger }\gamma _{\mu \nu }\Psi _{+},  \label{Seiberg-Witten2real}
\end{eqnarray}%
where $\Omega _{\mu }=\frac{1}{2}\Omega _{\mu ab}I_{ab}$ is the $SO(4)$
gauge potential, with $\Omega _{\mu }$ anti-Hermitian and $\Omega _{\mu
ab}\in \mathbb{R}$. $F_{+\mu \nu }$ is the self-dual component of the
curvature. $\mathbf{D}^{+A}$ is the twisted Dirac operator, $\mathbf{D}%
^{+A}:\Gamma (S^{+}\otimes L)\rightarrow \Gamma (S^{-}\otimes L)$, where $L$
is the twisting line bundle and $S^{\pm }\otimes L$ respectively the twisted
self-dual and anti-self-dual sub-spaces. $\Psi _{+}$ is the self-dual
solution of $\mathbf{D}^{+A}$. Eq.(\ref{Seiberg-Witten2real}) is the
so-called monopole equation. Given $\Omega _{\mu }$, the $A_{\mu }$ and $%
\Psi _{+}$ are two unknowns for the set of Eqs.(\ref{Seiberg-Witten1real},%
\ref{Seiberg-Witten2real}).

It is easy to recognize the similarity between the two sets Eqs.(\ref%
{Seiberg-Witten1real},\ref{Seiberg-Witten2real}) and Eqs.(\ref%
{1stSeibWittEqn'},\ref{2ndSeibWittEqn}). Hence the above theoretical
results, as well as the coming discussion about instantons, might give rise
to a new method for the study of the SW equations.

\section{Topological Characteristics of Instantons}

When (\ref{eigenValLap1}) holds, (\ref{4-SpinCdec5}) presents the $SU(2)_{+}$
gauge potential $\varpi _{\mu }$. In following it will be shown that the
spinor structure of $\varpi _{\mu }$ contributes to the second Chern classes
of the $SU(2)_{+}$ sub-bundle.

From the Chern-Simons $3$-form constructed with $\varpi $, $\Omega _{+}=%
\frac{1}{8\pi ^{2}}Tr(\varpi \wedge d\varpi -\frac{2}{3}\varpi \wedge \varpi
\wedge \varpi )$ (see Ref.\cite{DuanLiuCTP1} and references therein), using (%
\ref{SU(2)decomp2}) one acquires the Pauli spinor structure of $\Omega _{+}$%
\begin{equation}
\Omega _{+}=\frac{1}{4\pi ^{2}}\hat{\Phi}^{\dagger }d\hat{\Phi}\wedge d\hat{%
\Phi}^{\dagger }\wedge d\hat{\Phi}.  \label{chnsimnew}
\end{equation}%
The second Chern class $C_{2+}$ is given by $C_{2+}=d\Omega _{+}=\rho
_{+}(x)d^{4}x$, with $\rho _{+}(x)$ the so-called Chern density. Then
\begin{equation}
C_{2+}=\frac{1}{4\pi ^{2}}d\hat{\Phi}^{\dagger }\wedge d\hat{\Phi}\wedge d%
\hat{\Phi}^{\dagger }\wedge d\hat{\Phi},  \label{2ndchnclsnew}
\end{equation}%
i.e. the Chern density $\rho _{+}(x)=-\frac{1}{4\pi ^{2}}\epsilon ^{\mu \nu
\lambda \rho }\partial _{\mu }\hat{\Phi}^{\dagger }\partial _{\nu }\hat{\Phi}%
\partial _{\lambda }\hat{\Phi}^{\dagger }\partial _{\rho }\hat{\Phi}.$ This
expression is consistent with the $SU(2)$ Chern density obtained in Ref.\cite%
{DuanLiuCTP1}.

In order to study the inner structure of $C_{2+}$, we employ the component
form of the Pauli spinor $\Phi =(%
\begin{array}{cc}
\phi ^{0}+i\phi ^{1}, & \phi ^{2}+i\phi ^{3}%
\end{array}%
)^{T}$ and introduce a unit vector $n^{A}=\frac{\phi ^{A}}{\left\Vert \phi
\right\Vert }$ with $n^{A}n^{A}=1$. Here $\phi ^{A}\in \mathbb{R}$ and $\Phi
^{\dagger }\Phi =\phi ^{A}\phi ^{A}=\left\Vert \phi \right\Vert ^{2}$, $%
A=0,1,2,3$. Obviously the zero-points of $\phi ^{A}$ are the singular points
of $n^{A}$. Define a normalized spinor%
\begin{equation}
\hat{\Phi}=\frac{1}{\sqrt{\Phi ^{\dagger }\Phi }}\Phi =\left(
\begin{array}{c}
n^{0}+in^{1} \\
n^{2}+in^{3}%
\end{array}%
\right) .
\end{equation}%
Then in terms of $n^{A}$ one has $\rho _{+}(x)=-\frac{1}{12\pi ^{2}}\epsilon
^{\mu \nu \lambda \rho }\epsilon _{ABCD}\partial _{\mu }n^{A}\partial _{\nu
}n^{B}\partial _{\lambda }n^{C}\partial _{\rho }n^{D}$, which can be given
by a $\delta $-function form
\begin{equation}
\rho _{+}(x)=-\delta ^{4}(\vec{\phi})D(\frac{\phi }{x}),  \label{Roudelta}
\end{equation}%
where $D(\phi /x)$ is a Jacobian, $\epsilon ^{ABCD}D(\frac{\phi }{x}%
)=\;\epsilon ^{\mu \nu \lambda \rho }\partial _{\mu }\phi ^{A}\partial _{\nu
}\phi ^{B}\partial _{\lambda }\phi ^{C}\partial _{\rho }\phi ^{D}.$ Eq.(\ref%
{Roudelta}) implies that $\rho _{+}(x)=0$ iff $\vec{\phi}\neq 0$, while $%
\rho _{+}(x)\neq 0$ iff $\vec{\phi}=0$. So it is sufficient to study the
zero-points of $\vec{\phi}$ to determine the non-zero solutions of $\rho
_{+}(x)$. Meanwhile, one should bear in mind that $\left\Vert \phi
\right\Vert =0$ corresponds to the zero-eigenvalue of Eq.(\ref{eigenValLap1}%
).

The implicit function theory \cite{ImplicitFunc} shows that under the
regular condition $D(\phi /x)\neq 0$, the general solutions of the
zero-point equations $\phi ^{A}(x^{1},x^{2},x^{3},x^{4})=0$ are expressed as
$N$ isolated points: $x^{\mu }=x_{j}^{\mu },\;j=1,\cdots ,N$. In the
neighborhood of the $N$ zero points, $\delta ^{4}(\vec{\phi})$ is expanded
as \cite{Schouten}: $\delta ^{4}(\vec{\phi})=\sum_{j=1}^{N}\left. \frac{%
\beta _{j}\delta ^{4}(x^{\mu }-x_{j}^{\mu })}{\left\vert D(\phi
/x)\right\vert }\right\vert _{x_{j}^{\mu }}.$ Here the positive integer $%
\beta _{j}$ is called the Hopf mapping index. It means topologically that
when a point $x$ covers the neighborhood of the $j$th zero-point $x_{j}$
once, the vector field $\phi ^{A}$ will cover the corresponding region in
the $\phi $-space for $\beta _{j}$ times. Then, $\rho _{+}(x)$ is
re-expressed as%
\begin{equation}
\rho _{+}(x)=-\sum_{j=1}^{N}\beta _{j}\eta _{j}\delta ^{4}(x-x_{j}),
\label{35}
\end{equation}%
where $\eta _{j}=\left. \frac{D(\phi /x)}{\left\vert D(\phi /x)\right\vert }%
\right\vert _{x_{j}}=sign[D(\frac{\phi }{x})]_{x_{j}}$ is introduced, called
the Brouwer mapping degree. (\ref{35}) shows that the Chern density is
non-zero only at the $N$ $4$-dimensional zero-points of $\phi ^{A}$. These
self-dual singular-points are regarded as instanton solutions in the $%
SU(2)_{+}$ sub-space, with their topological charges characterized by $\beta
_{j}$ and $\eta _{j}$.

Integrating the second Chern class yields the second Chern number:
\begin{equation}
c_{2+}=\int C_{2+}=\int \rho _{+}(x)d^{4}x=-\sum_{j=1}^{N}\beta _{j}\eta
_{j}.
\end{equation}%
Since the base manifold $\mathcal{M}$ is $4$-dimensional, the second Chern
class $C_{2+}$ is the top Chern class, i.e., the Euler class on $\mathcal{M}$%
. Hence the Euler characteristic, which is the index sum of zero-points of
the smooth vector field $\phi ^{A}$ on $\mathcal{M}$ (Poincar\'{e}-Hopf
theorem), is obtained by the Gauss-Bonnet theorem:
\begin{equation}
\chi _{+}(\mathcal{M})=\int C_{2+}=-\sum_{j=1}^{N}\beta _{j}\eta _{j}.
\end{equation}%
This result demonstrates that the topological invariant $\chi _{+}(\mathcal{M%
})$ arises from the contribution of the topological charges of instantons.

\section{Conclusion and Discussion}

In this paper we obtain the Dirac spinor decomposition (\ref%
{4-SO(4)finaldecomp7}), i.e. (\ref{decomp1}), for the $Spin^{c}(4)$ gauge
potential $\omega _{\mu }+iA_{\mu }$. Comparing it with the decomposition of
the $SU(2)$ potential $\varpi _{\mu }$ we find a $U(1)$ characterizing
equation (\ref{eigenValLap1}). An important case for this equation is that $%
\lambda $ takes constant values. Then (\ref{eigenValLap1}) becomes an
eigenvalue problem of the Laplacian operator, with the eigenvalue being the
vacuum expectation value of $\Psi $. This thus provides a possible method
--- the spectrum of the Laplacian operator --- to determine the spinor field
and twisting potential in the Seiberg-Witten equations. Moreover,
topological charges of instantons are expressed in terms of the
characteristic numbers $\beta _{j}$ and $\eta _{j}$.

Some remarks are in order. Firstly, in Sect.3 one has extracted (\ref%
{eigenValLap1}) from (\ref{4-SpinCdec5}). Another interesting choice is to
modify the $U(1)$ potential in (\ref{4-SpinCdec5}) as $A_{\mu }=\bar{A}_{\mu
}+\alpha _{\mu }$. Here $\bar{A}_{\mu }$ satisfies (\ref{eigenValLap1}),
while $\alpha _{\mu }$ leads to a $\delta $-function like field strength:
\begin{equation}
d\alpha \propto \delta ^{2}(\vec{\varphi}),
\end{equation}%
with $\vec{\varphi}$ being a vector defined from the spinor $\Phi $ in the
self-dual sub-space. A choice for $\alpha _{\mu }$ is the Wu-Yang potential
\cite{DuanLiuPRD}:
\begin{equation}
\alpha _{\mu }\propto \vec{e}_{1}\cdot \partial _{\mu }\vec{e}_{2},
\end{equation}%
where $(\vec{e}_{1},\vec{e}_{2},\vec{m})$ is an orthogonal frame with $\vec{m%
}=\frac{\Phi ^{\dagger }\vec{\sigma}\Phi }{\Phi ^{\dagger }\Phi }$ defined
as a $3$-dimensional unit vector in the self-dual sub-space. Thus, $\alpha
_{\mu }$ will contribute to the LHS of (\ref{4-SpinCdec5}) a Chern-class
term containing a $2$-dimensional $\delta $-function, which indicates the
impact of the $2$-dimensional zero-points of $(\vec{e}_{1},\vec{e}_{2})$. $%
\alpha _{\mu }$ will also contribute to the RHS of (\ref{4-SpinCdec5}) a
Schwinger term related to a central extension, which describes the $3$%
-dimensional zero-points of $\vec{m}$.

Secondly, it has been pointed out that $\omega _{\mu }$ below (\ref%
{4-genDirac3}) is Hermitian but $\Omega _{\mu }$ in (\ref%
{Seiberg-Witten1real}) is anti-Hermitian. The physical meaning of imaginary
gauge potentials should be investigated \cite{ImPotent1}. Another
consideration for this problem is the following. Notice that in (\ref%
{4-genDirac3}) $\frac{1}{4}\bar{\omega}_{a}$, $\frac{1}{4}\tilde{\omega}_{a}$
and $iA_{a}$ are all purely imaginary potentials, i.e., they are in the
equal situation. Therefore one can regard that in (\ref{Seiberg-Witten1real}%
) an imaginary $SO(4)$ gauge potential $\Lambda _{\mu }$ can be introduced
to deliver the twisting action of the potential $iA_{a}$ exerted on $\Omega
_{\mu }$.

\emph{Acknowledgment:} The author X.L. is indebted to Prof. A.J. Bracken for
the discussion on Clifford algebra, and to M.C. Hong and Y. Zheng for their
help on the eigenvalue problem of the Laplacian operator. X.L. was
financially supported by the IPRS and UQGSS scholarships and the GSRTA award
of the University of Queensland. Y.S.D was supported by the National Natural
Science Foundation of P.R. China. W.L.Y. and Y.Z.Z. were supported by the
Australian Research Council.

\end{document}